\documentclass[aps,prb,reprint,floatfix,citeautoscript,longbibliography,superscriptaddress]{revtex4-1}
\usepackage{graphicx}
\usepackage{bm,amsmath,amssymb,mathrsfs,dcolumn}
\usepackage[usenames]{color}
\usepackage{subfigure}
\usepackage{ulem}
\begin{document}
\title{Skyrmions and multi-sublattice helical states in a frustrated chiral magnet}

\author{H. Y. Yuan}
\email[Corresponding author: ]{huaiyangyuan@gmail.com}
\affiliation{Institut f\"{u}r Physik, Johannes Gutenberg-Universit\"{a}t Mainz, 55099 Mainz, Germany}
\author{O. Gomonay}
\affiliation{Institut f\"{u}r Physik, Johannes Gutenberg-Universit\"{a}t Mainz, 55099 Mainz, Germany}
\affiliation{National Technical University of Ukraine ``KPI", 03056 Kyiv, Ukraine}
\author{Mathias Kl\"{a}ui}
\affiliation{Institut f\"{u}r Physik, Johannes Gutenberg-Universit\"{a}t Mainz, 55099 Mainz, Germany}
\date{\today}

\begin{abstract}
We investigate the existence and stability of skyrmions in a frustrated
chiral ferromagnet by considering the competition between ferromagnetic (FM)
nearest-neighbour (NN) interaction ($J_1$) and antiferromagnetic (AFM)
next-nearest-neighbour (NNN) interaction ($J_2$). Contrary to the
general wisdom that long-range ferromagnetic order is not energy preferable
under frustration, the skyrmion lattice not only exists but is even stable
for a large field range when $J_2 \leq J_1$ compared with frustration-free systems.
We defend that the enlargement of stability window of skyrmions is
a consequence of the reduced effective exchange interaction caused by the
frustration. A multi-sublattice helical state is found below the
skyrmion phase, which results from the competition between AFM coupling
that favors a two-sublattice N\'{e}el state and the chiral interaction that prefers
a helix. As a byproduct, the hysteresis loop of the frustrated chiral system
shrinks as the magnetization goes to zero and then opens up again, known
as wasp-waist hysteresis loop. The critical field that separates the
narrow and wide part of the wasp-waist loop depends exponentially on the
strength of NNN coupling. By measuring the critical field, it is possible
to determine the strength of NNN coupling.
\end{abstract}

\maketitle

\section{Introduction}
Frustration in spin systems refers to competing interactions
that cannot be satisfied simultaneously and this phenomenon
has attracted significant scientific attention in last few decades
due to its unusual ordering properties of the ground states
\cite{Wannier1950, Stephenson1970, Marland1979, Balents2010, Binder1986}.
One type of frustration is geometric frustration, where the frustration
is caused by the special structure of the crystal lattice
\cite{Wannier1950, Chalker1992}. A well-known example is the antiferromagnetically
 coupled Ising spin model with $S=1/2$ on a two-dimensional (2D) triangle
lattice \cite{Wannier1950}. With two spins on the vertices aligned
antiparallel to each other, the third spin cannot find an preferable
orientation to gain exchange energy, which results in the absence of
long-range ferromagnetic order and high degeneracy of the ground states.
The other type of frustration comes from the competing magnetic exchange
interactions instead of the lattice itself
\cite{Villain1977, Vannimenus1977, Chandra1988, Tsirlin2010,Nath2008}.
Typical examples are 2D square lattices with competing nearest-neighbor
(NN) interaction $J_1$ and next-nearest-neighbor (NNN) interaction $J_2$
\cite{Tsirlin2010}. Depending on the sign of $J_1, J_2$ and the ratio of
$J_2/J_1$, the ground state of the system could be a columnar antiferromagnet (AFM)
(2-sublattice AFM), a N\'{e}el AFM (4-sublattice AFM), and a disordered
phase \cite{Shannon2006}. Generally, the existence of frustration would
suppress the formation of long-range ordering phase and give rise to
some short-range ordered phase such as spin glass and spin liquid
\cite{Binder1986,Balents2010}.

The skyrmion lattice, a long-range ordered vortex-like topological magnetic
structure, has been recently realized in chiral magnets
\cite{Bogdanov2001, Robler2006,Muhlbauer2009,Yu2010,Munzer2010,Yu2011,Heinze2011,Onose2012,Schulz2012,Mohseni2013,
Buhrandt2013, Fert2013, Iwasaki2013, Park2014, Woo2016} with Dzyaloshinskii-Moriya (DM) interaction
\cite{Dzy1957, Moriya1960}.
Each skyrmion consists of spins that point in all directions
wrapping a unit sphere. Skyrmion can be represented by a topological charge
$q = 1/4\pi \int  dx dy \mathbf{S}\cdot (\partial_x \mathbf{S}
\times \partial_y \mathbf{S})=\pm 1$, where $\mathbf{S}$ is the unit
vector of spin and the Cartesian coordinates $x-y$ lie in the film plane.
Then skyrmion lattice could be viewed as regularly patterned topological
charges, whose stability is topologically protected. An interesting question
is how topologically non-trivial structures like skyrmions respond to the
frustration in a system? A pioneering work studied the isotropic
Heisenberg model on a triangular lattice and found that a skyrmion lattice
and multi-q states could exist at non-zero temperatures even in the absence
of DM interaction \cite{Okubo2012}. A followed work focused
on the anisotropic frustrated spin model and showed that multiply periodic states even
exist at zero temperature \cite{Leonov2015}. However, yet to be known is how the frustration
influences the existence and energy-preferable window (EPW) of a skyrmion
lattice stabilized by DM interaction. That is the motivation of the current work.

In this article, we focus on a frustrated square lattice with DM interaction.
The frustration is introduced through the competing between NN FM interaction
and NNN AFM interaction. The energy-preferable window of the skyrmion phase as
well as hysteresis loop of the frustrated system are studied and discussed
in detail. The paper is organized as follows. In Sec II, the model and
numerical method are introduced. In Sec III, the phase diagram of skyrmions
and wasp-waist hysteresis loop are presented, together with detailed
analysis and discussions. The conclusions are given in Sec IV.

\section{Model and Method}
We consider a magnetic thin film with square lattice where $x-$, $y-$ and $z-$
axes are along the length, width and thickness directions, respectively.
The lateral dimensions of the film are are $L \times L$.
The Hamiltonian of the system is

\begin{equation}
\begin{aligned}
H &= -J_1 \sum_{<i,j>} \mathbf{S}_i \cdot \mathbf{S}_j + J_2\sum_{<<i,j>>}
\mathbf{S}_i\cdot \mathbf{S}_j \\
&+ \sum_{<i,j>} \mathbf{D} \cdot \mathbf{S}_i \times \mathbf{S}_j
-H \sum_{i} \mathbf{S}_i^z,
\end{aligned}
\end{equation}
where $\mathbf{S}_i$ labels the classical spin orientation at site $i$,
the first and second sums are taken over all NN and NNN pairs, respectively
and $J_1$, $J_2$ refer to corresponding exchange interaction strength.
The third term is DM interaction which results in the formation of
skyrmions in the lattice and $\mathbf{D}$ is the DM vector.
This term is one type of Lifshitz invariant that
stabilizes a skyrmion state \cite{Izyumov1984, Bogdanov1989, Leonov2016}. The fourth
term is Zeeman energy, where the external field $\mathbf{H}$ is along the
thickness direction ($+z$). A single spin flip
Monte Carlo (MC) method \cite{Metropolis1953} is used to simulate the ground states.
To mimic an infinite system, periodic boundary condition in the $x$ and
$y$ directions are used. A typical simulation begins with a completely
random state at a sufficient high temperature $T=10\ J_1/k_B$, where
$k_B$ is Boltzmann constant, and then the system is annealed to target
temperature after a reasonable number of temperature steps.
At each temperature, $10^4$ MC steps are taken before measurements
are performed. If not stated differently, the grid size $L=32$ and NN
coupling is set to $J_1 = 1.0$. It corresponds to exchange integral
in the order of 10 meV \cite{Kittel1996}.

\section{Results and Discussions}
In Sec A, we will first illustrate our simulation results including
the phase diagram in $H-J_2$ plane, the comparison of energy between
ferromagnetic (FM), skyrmion (SkX) and helical states (H), and then
provide both qualitative and quantitative analyses of the results.
In Sec B, we present the results on hysteresis loop and then discuss
the results in detail.

\subsection{Phase diagram}
\begin{figure}
\centering
\includegraphics[width=0.45\textwidth]{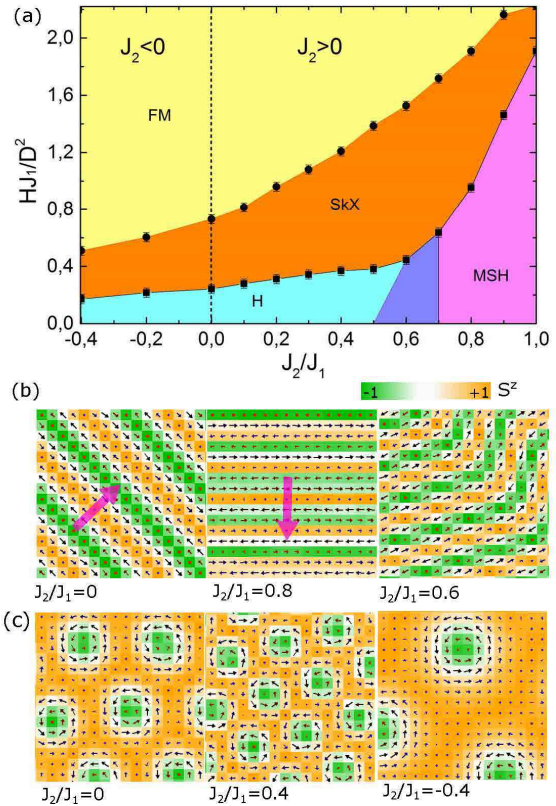}\\
\caption{(Color online) (a) The $HJ_1/D^2-J_2/J_1$ phase diagram when
NNN coupling is included. $T=6.6 \times 10^{-3} J_1/\mathrm{k_B}$.
The color represents various phases including yellow (FM), orange (SkX),
light blue (H), dark blue (Transition) and pink (MSH).
The dashed line refers to $J_2=0$.
(b) Ground states of the system for $J_2/J_1 = 0,\ 0.8,\ 0.6$ under
zero field. The pink arrows indicate the direction of period.
(c) Snapshot of skyrmions for $J_2/J_1 = 0,\ 0.4, -0.4$ under field
$HJ_1/D^2 = 0.45$. The color codes $S_z$ from -1 (green), 0 (white)
to +1 (orange) as indicated in the color bar.}
\label{fig1}
\end{figure}

Figure \ref{fig1}a shows the phase diagram in the $HJ_1/D^2-J_2/J_1$
plane. When NNN coupling is absent i.e. $J_2=0$, the ground state is a
helical state for $HJ_1/D^2<0.24$ as shown in the left panel of
Fig. \ref{fig1}b. The chirality of the helix is counterclockwise,
which is uniquely determined by the sign of the DM interaction \cite{Yuan2016}.
For $ 0.24\leq HJ_1/D^2 < 0.73$, the ground state is a triangular
skyrmion lattice as shown in the left panel of Fig. \ref{fig1}c.
It should be mentioned that the number of skyrmions decreases
and the hexagonal structure of skyrmion lattice becomes irregular
when the reduced field is close to 0.73. This observation may be related
strip-out instability of skyrmion lattice. \cite{Leonov2016}
As the field increases further, the ground state is a ferromagnetic
single domain, where all the spins align in the $+z$ direction.
Now, the antiferromagnetic NNN coupling ($J_2>0$) is turned on.
As $J_2/J_1$ increases, the EPW of skyrmion phase first enlarges
and then shrinks as shown in the orange region of Fig. \ref{fig1}a.
Up to $J_2/J_1 =0.9$, the EPW is wider than the case of $J_2=0$
and the maximum enlargement ratio is 2.2 around $J_2/J_1 =0.6$.
As a comparison, we simulate the case when NNN coupling is
ferromagnetic ($J_2 <0$) and find that stability region of skyrmion shrinks.
The phase above the upper boundary of skyrmion phase (yellow) is always
a FM state while the phase below depends on the strength of $J_2/J_1$.
For $J_2/J_1 \leq 0.5$, the helical state is stable. As $J_2/J_1 >0.7 $,
a multi-sublattice helical (MSH) phase as shown in the middle panel
of Fig. \ref{fig1}b, becomes stable. In the intermediate regime,
it is the transition state as shown in the right panel of Fig. \ref{fig1}b.

\begin{figure}
\centering
\includegraphics[width=0.45\textwidth]{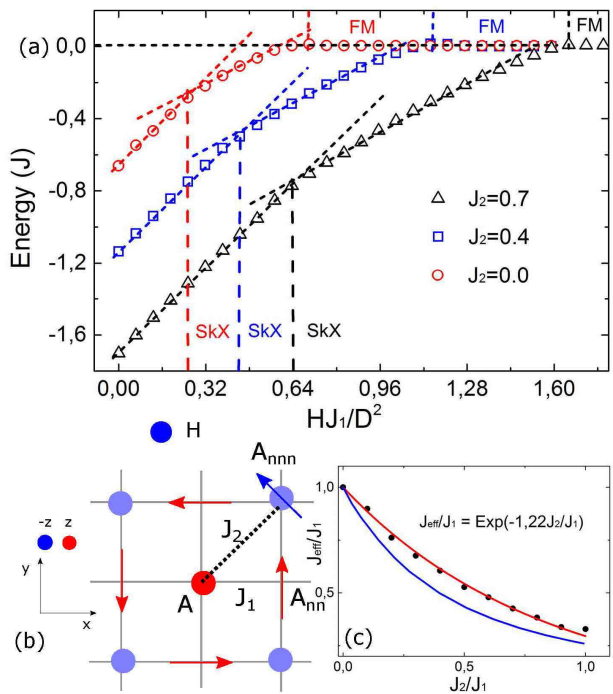}\\
\caption{(Color online) (a) Energy of the system as a function of
external field for $J_2 = 0.0$ (red circles), $0.4$ (blue squares)
and $0.7$ (black triangles). The energy is biased by the energy
of a FM state. The dashed lines indicate the corresponding phase boundaries.
(b) Scheme of a site A and its neighbours $\mathrm{A}_{\mathrm{nn}},
\mathrm{A}_{\mathrm{nnn}}$ on a square lattice.
(c) Effective NN exchange $J_{\mathrm{eff}}$ as a function of NNN exchange
strength. The red line is the fitting curve using relation
$J_{\mathrm{eff}}/J_1 = \mathrm{Exp}(-1.22 J_2/J_1)$. The blue line
is the theoretical prediction.}
\label{fig2}
\end{figure}

A consistency check is performed by comparing the energy of skyrmions
and helical states as shown in Fig. \ref{fig2}a. The energy curve shows that
the normal helical or multi-sublattice helical state has the lowest
energy when the field is small, regardless of the magnitude of $J_2$
(red circles, blue squares and black triangles for $J_2 = 0.0, 0.4, 0.7$,
respectively). the FM state is energy preferable in the large field limit.
In the intermediate field regime, the skyrmion phase is energy preferable
and the EPW of skyrmions (the horizontal window between two dashed
lines with the same color) increases as NNN coupling increases from
0 to 0.7. As a short summary, both the phase diagram and energy
comparison between various states explicitly show that a skyrmion
lattice could exist and the energy-preferable window of skyrmions
becomes even larger for $0 < J_2/J_1<1.0$. Moreover,
the energy changes are continuous while the first order
derivative of energy with respect to field is not continuous at the transition fields,
as shown by the dashed lines connecting the discrete symbols in Fig. \ref{fig2}a.
This indicates that the transition between the helical state and the skyrmion,
the skyrmion and the FM state may be both the first order transitions.

Now we provide a qualitative understanding on the phase diagram and energy
landscape. Firstly, as shown in Fig. \ref{fig2}b, one spin A has four
nearest spins $A_{nn}$ and another four NNN spins $A_{nnn}$. The introduction
of an antiferromagnetic NNN coupling would prefer the spin $A_{nnn}$ to
tilt antiparallel to the central spin $A$. Even though this trend would
be suppressed by the ferromagnetic NN coupling of $A_{nn}$, it would
weaken the NN coupling strength effectively. As a result, the effect
of the DM interaction would be more pronounced and make the skyrmion more
energetically favorable. On the other hand, the reduction of effective
exchange coupling would lead to a reduction of skyrmion size \cite{JHHan2010},
which is consistent with our observation as shown in Fig. \ref{fig1}c.
If it is assumed that the phase boundary between SkX and FM in Fig. \ref{fig1}a
still follows the theoretical relation
$H_cJ_{\mathrm{eff}}/D^2$ = 0.73 \cite{JHHan2010} in a frustration-free
system, the effective coupling $J_{\mathrm{eff}}$ as a function of $J_2$
could be extracted as shown by the black dots in Fig. \ref{fig2}c.
An exponential fitting $J_{\mathrm{eff}}/J_1 = \mathrm{Exp}(-1.22 J_2/J_1)$
 describes the NNN coupling strength $J_2$ dependence of $J_{\mathrm{eff}}$ well.

\begin{figure}
\centering
\includegraphics[width=0.45\textwidth]{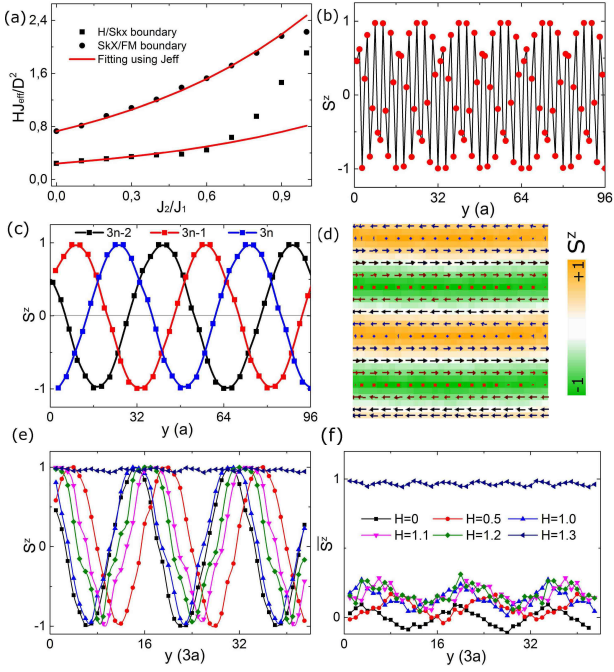}\\
\caption{(Color online) (a) Calculated phase boundary using effective NN
coupling $J_{\mathrm{eff}}$ (red line) for H/SkX ($0.24HJ_{\mathrm{eff}}/D^2$)
and SkX/FM ($0.73HJ_{\mathrm{eff}}/D^2$).
(b) $S^z$ as a function of the spin position ($y$), where $a$ is lattice constant.
(c) Re-plot of $S^z-y$ when the sites is classified into three classes
$i=3n-2, 3n-1, 3n$, where $n$ is a positive integer. $y= ia$,
where $a$ is lattice constant. Black, red, blue
squares represent $3n-2, 3n-1, 3n$, respectively. (d) A typical spin
configuration of the $3n-2$ sites. (e) $S^z-y$ for $3n-2$ sites at
various fields. (f) The average magnetization of the $3 n-2,3n-1$ and
$3n$ sites $\overline{S^z}$ as a function of unit cell coordinates ($y$)
for various fields. $\overline{S^z}=(S^z_{3n-2} + S^z_{3n-1} + S^z_{3n})/3$.}
\label{fig3}
\end{figure}

However, the effective ferromagnetic coupling cannot capture the full
physics in a frustrated system. As shown in Fig. 3a, the effective exchange cannot
fully explain the phase boundary between the MSH state and the SkX state
for $J_2/J_1>0.5$. This may be attributed to the speciality of the
MSH state compared with the conventional helical state.
Next, we discuss the properties of MSH in detail.
For the typical example shown in the middle panel of Fig. \ref{fig1}b,
Figure 3b shows the $z$ component of spin ($S_z$) as a function
of spin position in $y$ direction. Although it shows a periodic behavior
with a period of $L_H=32a$ ($a$ is the lattice constant),
it is difficult to recognize the multi-sublattice
chiral structure from this plot directly. Here we introduce a unit cell
containing three sublattices with site number $3n-2$, $3n-1$, $3n$, respectively,
where $n$ is a positive integer. Figure 3c shows the space variation of spins
of the three sublattices, respectively. A cosine/sine curve with period
$L_H=48a$ is observed for all the three sublattices and a constant phase lag
exists for adjacent sublattices. If we only plot the spin configuration
of one sublattice, it shows a regular helical structure as shown in
Fig. 3d for $3n-2$ sites. The chirality of the helix in $+y$ direction
is counterclockwise, which is the same as the helical states
at small $J_2$. This multi-sublattice helical state is the competing
result between DM interaction and exchange coupling.
Specifically, when $J_2/J_1<0.5$, the NNN coupling is weak, hence
ferromagnetic NN coupling and DM interaction dominate the other
interactions in the system and the ground state is a helical state.
When $J_2/J_1 > 0.5$, the AFM NNN coupling begins to surpass the NN
coupling and compete with the DM interaction. Without DM interaction,
the ground state is a 2-sublattice
AFM state \cite{Tsirlin2010}, where spins in a row/column would align
ferromagnetically with each other and the spins in a column/row align
antiferromagnetically. Once DM interaction appears, it will prefer a chiral rotation
of spins in a particular direction. As a result of this balance, the spins
keep their ferromagnetic order in the $x$ direction and change the 2-sublattice
AFM to a three sublattices helical state in $y$ direction.
The multi-sublattice state is very stable as it gains from both the AFM
interaction and DM interaction. It should be pointed out that the multi-sublattice
ordering is also possible in the $x$ direction, since there is no anisotropy
in our 2D square lattice.
Figure 3e shows the $S_z-y$ for the $3n-2$ sublattice as we increases
the fields from 0 to 1.3 at the phase boundary. It shows that the
sublattice chiral state is stable as the field increases up to 1.2.
This special stability widens the window of multi-sublattice state
and shrinks the window of skyrmions, as shown in the phase diagram of
Fig. \ref{fig1}a. The phase difference of the $S^z-y$ curves between various
fields in Fig. \ref{fig3}e is randomly distributed, which is related to the stochastic
nature of spin flip in the annealing process. Nevertheless, as the field increases,
the phase difference of the three sites within a unit cell keeps adjusting
to increase the average magnetization to gain Zeeman energy, as shown in Fig. \ref{fig3}f.

\begin{figure}
\centering
\includegraphics[width=0.45\textwidth]{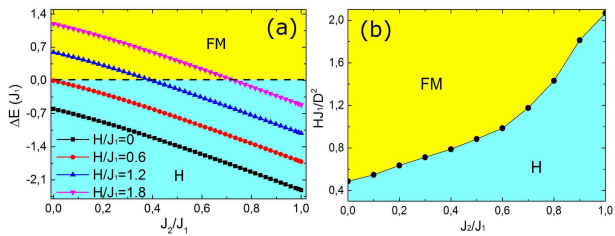}\\
\caption{(Color online) (a) Energy density difference between a helical
state and a FM state calculated from 1D frustrated model under fields
$H/J_1$ =0 (black square), 0.6 (red dots), 1.2 (blue up-triangles),
1.8 (pink down-triangles). The light blue/yellow region refers
to the region with FM and helix as ground states, respectively.
(b) The calculated phase diagram in $HJ_1/D^2-J_2/J_1$ plane.
The yellow and light blue color represent FM and H states, respectively.}
\label{fig4}
\end{figure}

To justify the results theoretically, we model a frustrated spin chain
and study the influence of frustration on the chiral states. The Hamiltonian is
\begin{equation*}
\begin{aligned}
H=\sum_i \left (-J_1 \mathbf{S}_{i} \cdot \mathbf{S}_{i+1}
+ J_2 \mathbf{S}_i\cdot \mathbf{S}_{i+2} \right. \\
\left . -D e_y \cdot \mathbf{S}_i \times \mathbf{S}_{i+2}
-H  \mathbf{S}_i^z \right ).
\end{aligned}
\end{equation*}
The energy of FM is $E_{\mathrm{FM}}/J_1=-1+J_2/J_1 - H/J_1$.
For the helical state, it is assumed that the spin rotates in the $xz$ plane
such that $S_n=(\sin nqa, 0, \cos nqa)$, where $a$ is lattice constant,
and the wavevector $q=2 \pi /L_H$, where $L_H$ is the period of the helix.
Substituting this trial function $S_n$ into the Hamiltonian and minimizing
the total energy gives a quartic polynomial equation,

$$Dx^4 +(2J_1 + 4J_2) x^3 + (2J_1-4J_2)x - D =0$$
where $x=\tan (qa/2)$. For $J_2=0$, the solution is $J_1 \tan (qa) = D$,
which is consistent with the literature \cite{Yi2009}. For $J_2 \neq 0$,
the quartic solution has a positive real root ($q_0$), which indicates
the existence of a helical structure. Substituting this root ($q_0$)
back to the Hamiltonian, we could calculate the energy of the helix.
Figure \ref{fig4}a shows the energy difference between the helical
state and FM state as a function of $J_2/J_1$. For $H/J_1<0.6$,
the energy of helix is always smaller than the energy of FM state.
For $H/J_1 > 0.6$, the helix is energetically preferred in the larger
$J_2/J_1$ regime while FM is energetically preferred in the smaller
$J_2/J_1$ regime. The phase diagram in the $J_2/J_1 -HJ_1/D^2$
plane is shown in Fig. \ref{fig4}b. By fitting the phase boundary
between helix and FM state using $HJ_{\mathrm{eff}}/D^2$, we could
extract the effective exchange as a function of $J_2/J_1$ in the frustrated
system, as shown in the blue line of Fig. \ref{fig2}c. The theoretical curve
captures the trend of numerical results (black dots) in the 2D square lattice.

\subsection{Hysteresis loop}
\begin{figure}
\centering
\includegraphics[width=0.45\textwidth]{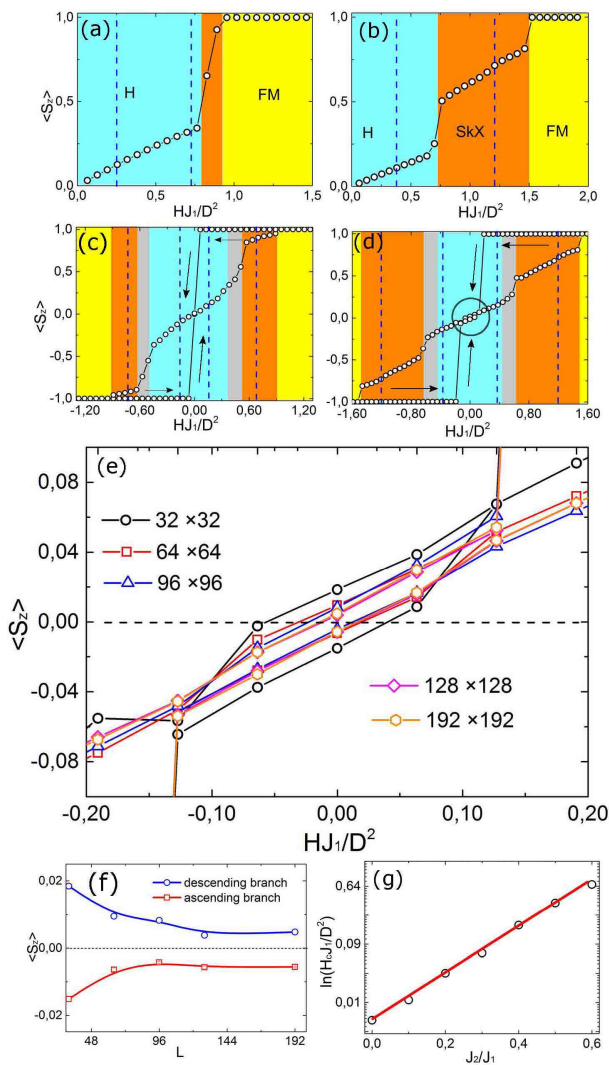}\\
\caption{(Color online) First magnetization curve for $J_2=0$ (a)
and $J_2 =0.4$ (b), and hysteresis loop for $J_2/J_1 = 0.0$ (c) and
0.4 (d). $<S_z>$ is defined as $<S_z> =1/L^2\sum_i S^z_i$.
Blue: helix, orange: skyrmion, yellow: FM state, gray: mixing of helix
and skyrmion.  (e) is the enlarged figure of the circle in (d).
$T=6.6 \times 10^{-3}$ $\mathrm{J_1/k_B}$, $L=32$.
(e) Size effect of the hysteresis loop at small fields for
$L=32$ (black circles), 64 (red squares), 96 (blue triangles)),
128 (pink diamonds), 192 (yellow hexagons), respectively.
(f) $<S_z>$ as a function of sample size for descending branch
(blue circles) and ascending branch (red squares), respectively.
(g) The critical field $H_c$ as a function of $J_2/J_1$.}
\label{fig5}
\end{figure}

In this section, we investigate the shape of hysteresis loop in a
frustrated system. The thermal effect is suppressed by considering a
system at sufficiently low temperature $T=6.67\times 10^{-3}\ \mathrm{J_1/k_B}$.
Figure \ref{fig5}a,b shows the first magnetization curve. The average magnetization
is defined as $<S_z> =1/L^2\sum_i S^z_i$. To simulate the
first magnetization, we start from a helical state obtained from an annealing
process under $H=0$ and then increase the field in the step of $0.1J_1/D^2$.
For $J_2=0$, the helical phase is stable up to $HJ_1/D^2=0.76$ and
skyrmion phase only exists in a narrow window for $0.76<HJ_1/D^2<0.95$,
as shown in Fig. \ref{fig3}a. For $J_2=0.4$, skyrmion is stabile at
the range $0.70<HJ_1/D^2<1.52$, which is significantly larger than the
window at $J_2=0$. For both $J_2=0$ and $J_2 =0.4$, the existence regime of the skyrmion
phase is different from the phase diagram shown in Fig. \ref{fig1}a, which
is due to the hysteretic nature of a magnetic system \cite{Buhrandt2013}.
The final spin configuration depends sensitively on the evolution
history, even though the initial states are the same.
The annealing procedure is more efficient to reach the real
global minimum states of the system because sufficient energy is provided
to the spins to overcome the energy barrier between skyrmion and H/FM.
Under a fixed low temperature ($6.67\times 10^{-3}\ J_1 \mathrm{/k_B}$),
the thermal energy is too small for the system to overcome the energy
barrier to reach a skyrmion state unless sufficient large field is
applied such that the energy barrier is substantially lowered.

Figure \ref{fig5}c and \ref{fig5}d show the hysteresis loop for $J_2=0$
and $J_2 = 0.4$, respectively. To simulate the hysteresis loop, we start
from a FM state obtained by annealing procedures at high fields and then
decrease the fields in the step of $0.1J_1/D^2$ to obtain the descending
branch and then reverse the direction of fields to get the ascending branch.
Both the hysteresis loops in Fig. \ref{fig5}c and \ref{fig5}d take on
a wasp-waist shape, where the loop shrinks as magnetization goes to zero
and then opens up again. The field that separates the wide and narrow
part of the hysteresis loop is denoted as the critical field ($H_c$).
The critical field is $H_c=0$ for $J_2=0$ while it is at
$H_c = 0.13 D^2/J_1$ for $J_2 = 0.4$, as shown in Fig. \ref{fig5}c
and \ref{fig5}d. To see clearly the loop in the small magnetization regime
and understand the role of finite size effect in the simulations, we vary the system
size as $L=32, 64, 96, 128, 192$ and plot all the loops in Fig. \ref{fig5}e.
For $L=32$ (black dots), a quasi-closed loop exists, and it
disappears as $L$ increases. The $L$ dependence of the magnetization
for the ascending branch and descending branch at zero field is plotted in
Fig. \ref{fig5}f. The gap between the two branches decreases gradually
to a constant. It is expected that the loop at smaller fields would
result in two closely parallel branches as $L \rightarrow \infty$.

To understand the wasp-waist hysteresis, we refer to the scheme in
Fig. \ref{fig2}b. As we reduce the fields slowly from a high field
value, the system first stabilizes at a FM state, where spins at
$A$ and its neighbors $A_{nn}$ and $A_{nnn}$ (almost) align in the
$+z$ direction. Similar to the argument for the first magnetization curve,
even though the skyrmion/helical states are the global minimum states
for $HJ_1/D^2<0.76$, however the FM state ($q=0$) and skyrmions
($q=\pm 1$) are topologically different, where the transformation
between them needs
to overcome a finite energy barrier. As the temperature is low, the
degree of freedom for the spins frozen and the system always tends to evolve to
a lower energy state, which makes it difficult to climb across the barrier.
The system finally evolves to a helical state ($q=0$) at zero field.
When antiferromagnetic NNN coupling
is present, the spins between $A_{nn}$ and $A$ will tend to align
antiparallel to each other, which will destabilize the FM state and
lower the energy barrier between FM and helical states. That's why
the critical field becomes finite. According to
this argument, the larger the NNN coupling ($J_2$), the lower the barrier
between the FM and helical states, hence the larger the
critical field. This is demonstrated by the simulations as shown in
the inset of Fig. \ref{fig5}g. The critical field increases almost
exponentially with the increase of $J_2$. This may provide a method
to measure the strength of NNN coupling in a ferromagnetic film by
measuring the critical field in the corresponding wasp-waist hysteresis
loop.

Lastly, we would like to mention that a similar wasp-waist hysteresis
loop was experimentally observed in Co/Pt multi-layer structures with
low disorder \cite{Pierce2007}. At the critical field that magnetization
decreases significantly, a labyrinth-like magnetic pattern was observed.
Another system that allows for the wasp-waist loop is a chiral system with in-plane
easy-axis anisotropy \cite{Kwon2012}. The critical field of the hysteresis
loop depends on the in-plane anisotropy strength. One should be careful to
clarify the source of the critical field when NNN coupling and easy-axis
anisotropy are possible to coexist in a system. Moreover, it is
known that the exchange coupling of Fe could be both ferromagnetic and
antiferromagnetic which depends on the substrates' d-band filling
\cite{Hardrat2009, Rozsa2015, Rozsa2016prb}. Recently, \textit{ab initio}
calculation showed that ferromagnetic NN and AFM NNN interaction could coexist
in multi-layer $\mathrm{Pt}_{1-x}\mathrm{Ir}_x$/ Fe/Pd structures \cite{Rozsa2016},
where our simulation results are promising and remain to be verified experimentally.

\section{Conclusions}
In conclusion, we have studied the existence and the energy-preferable
window of skyrmions in a frustrated magnet. The AFM NNN interaction
diminishes the FM NN interaction to a certain degree and the DM interaction
becomes more pronounced, hence the skyrmions could be stabilized in
an even larger window compared to a system without frustration.
The theory based on the effective ferromagnetic exchange could
quantitatively capture the phase boundary among skyrmions, FM states
and the helix, except for the skyrmion/helix boundary when the AFM coupling
is larger than half of the FM coupling. In this regime, a multi-sublattice
state instead of a conventional helical state exists below the skyrmion phase.
The multi-sublattice state gains both AFM and DM energy and is energetically
preferable up to high fields. Moreover, the hysteresis loop
of the frustrated system takes on wasp-waist shape and the critical
field at which the loop shrinks depends on the strength of NNN
coupling. By measuring the critical field, it is possible to
determine the strength of NNN coupling.

\acknowledgments
We acknowledge the support from Deutsche Forschungsgemeinschaft (DFG) via the
Transregional Collaborative Research Center (SFB/TRR) 173
``Spin+X - Spin its collective environment" and
 the ERC Synergy Grant SC2 (No. 610115).

\end{document}